\title{A Kalman Filter Algorithm with Process Noise Covariance Update}

\author{
        Krishan Kumar Gola, Shaunak Sen\\
        Department of Electrical Engineering\\
        Indian Institute of Technology Delhi\\
        Hauz Khas, New Delhi 110016, INDIA\\
}

\date{}

\documentclass[12pt]{article}

\usepackage[paper=letterpaper, dvips, top=1.4cm, left=1.75cm, right=1.75cm, foot=1cm, bottom=1.5cm]{geometry}
\usepackage{setspace}
\usepackage[paper=letterpaper]{geometry}
\usepackage{graphicx, subfigure}
\usepackage{url}
\usepackage{amsmath, amsthm} 
\usepackage{amssymb}  
\usepackage{breqn}
    
\newtheorem{thm}{Theorem}

\begin{document}
\maketitle
\noindent
\textbf{Abstract.} 
Stochastic models in biomolecular contexts can have a state-dependent process noise covariance.
The choice of the process noise covariance is an important parameter in the design of a Kalman Filter for state estimation and the theoretical guarantees of updating the process noise covariance as the state estimate changes are unclear.
Here we investigated this issue using the Minimum Mean Square Error estimator framework and an interpretation of the Kalman Filter as minimizing a weighted least squares cost using Newton's method. 
We found that a Kalman Filter-like algorithm with a process noise covariance update is the best linear unbiased estimator for a class of systems with linear process dynamics and a square root-dependence of the process noise covariance on the state.
We proved the result for discrete-time system dynamics and then extended it to continuous-time dynamics using a limiting procedure.
For nonlinear dynamics with a general dependence of process noise covariance on the state, we showed that this algorithm minimizes a quadratic approximation to a least squares cost weighted by the noise covariance.
The algorithm is illustrated with an example.
\\\\

\section{Introduction}
Mathematical models of biomolecular systems may have an explicit state and parameter dependence in the process noise term,
\begin{eqnarray}
dx_i =  \sum_{j=1}^M \nu_{ij}a_j(x)dt + \sum_{j=1}^M\nu_{ij}\sqrt{a_j(x)}N_j(t)\sqrt{dt}, 
\end{eqnarray}
where $i = 1, 2, \ldots, n$, $x = [x_1, x_2, \ldots x_n]'$ is a state vector of the biomolecular concentration levels, $a_j$ represent different biochemical reactions, $\nu_{ij}$ is the change in biomolecular species $x_i$ due to the reaction $a_j$, and $N_j(t)$ are independent Gaussian white noise processes with zero mean and unit variance~\cite{del_vecchio_biomolecular_2015}.
These dynamics constitute a Chemical Langevin Equation for the particular reaction scheme.
Depending on whether the output equation is continuous or discrete, the overall dynamics may be viewed as continuous or continuous-discrete, respectively~\cite{jazwinski_stochastic_2007}.
Similar models arise in other contexts such as in finance~\cite{kloeden_numerical_1992}.
The estimation of states and parameters from output measurements is an important problem in these contexts.

The Kalman Filter is a standard method for state estimation, and the filter equations can be obtained either from a probabilistic approach that minimizes the expected value of a mean square error or from a statistical approach that minimizes a weighted least square cost function~\cite{jazwinski_stochastic_2007},\cite{humpherys_fresh_2012}.
While the Kalman Filter is an optimal filter for linear systems driven by white Gaussian noise, it is often extended to other nonlinear non-Gaussian contexts with good results in practise but no theoretical guarantees~\cite{anderson_optimal_1979}.
The general nonlinear estimation problem may be formulated as a Kushner equation, but the solutions typically involve approximations~\cite{jazwinski_stochastic_2007}.
Therefore, extended variants of the Kalman Filter are typically used for estimation in nonlinear contexts, including in the above contexts~\cite{lillacci_parameter_2010}.
The choice of the process noise covariance is an important design parameter in the design of a Kalman Filter.
This is often chosen heuristically, although there are adaptive algorithms as well~\cite{anderson_optimal_1979},\cite{speyer_stochastic_2008}.
The impact of the state dependence of the process noise covariance on the estimation algorithm is relatively less explored.

We had previously asked whether and how the knowledge of the process noise term may be used to augment state and parameter estimation in a Kalman Filter~\cite{dey_kalman_2019}.
Our numerical study showed that, in comparison with the standard Kalman Filter with a fixed process noise covariance, a Kalman Filter variant where process noise covariance is updated at each time-step based on the new estimate may have encouraging properties including convergence of the state and parameter estimates, an innovation sequence that is white, and an optimal balance of mean square estimate error and convergence time.
This variation, motivated heuristically by the system properties, may be viewed as a direct consequence of linearising the system model, including the process noise term, at each time step, as an Extended Kalman Filter.
However, this relies on ``closeness'' of the nonlinear system with its linearized approximation and a complete analytical understanding of the improved performance is unclear.

Here we investigated the theoretical basis for the improved performance of such a Kalman filter algorithm.
We showed that this Kalman Filter algorithm is the Best Linear Unbiased Estimator for linear system dynamics with process noise that depends on states in a square root-like fashion.
We proved this for discrete system dynamics and, using a limiting procedure, for continuous system dynamics.
For nonlinear system dynamics and associated process noise covariances, we showed that this algorithm minimizes a quadratic approximation to a noise covariance-weighted least squares cost.

The result was illustrated through an example.

\section{Theoretical Results}

For simplicity, we first considered a discrete-time version.
The following theorem provided a theoretical justification for the Kalman Filter variant where the process noise covariance is updated at each time step.
We showed that the Kalman Filter variant is the Best Linear Unbiased Estimator in that it minimized the mean square error among all affine unbiased estimators.
We used $\hat{x}_{k | k - 1}$ and $\hat{x}_{k | k}$ to denote the estimate of the state $x_k$ at times $k - 1$ and $k$, respectively.
The associated measurement covariances were denoted by the symbol $\Sigma$.

\begin{thm}
The Best Linear Unbiased Estimator for the discrete-time system
\begin{eqnarray}
x_{k+1} &=& A_0 + A_1x_k + G(x_k)v_k \\
y_k &=& Cx_k + w_k,
\end{eqnarray}
where $k = 0, 1, \ldots$, $x_k \in \mathbb{R}^n$, $y_k \in \mathbb{R}^m$, $\{v_k, k =1,\ldots\}$ is an $n$-vector, white noise sequence $v_k \sim (0,\Sigma_v)$ with $\Sigma_v$ diagonal, $w_k \sim (0, \Sigma_w)$, and $x_0$, $\{v_k\}$, and  $\{w_k\}$ are assumed independent, and $G(x_k) = diag[g_1(x_k), g_2(x_k), \ldots g_n(x_k)]$ is\\
Measurement update:
\begin{eqnarray}
\hat{x}_{k|k} =& \hat{x}_{k|k-1} + \Sigma_{k|k-1}C'(C\Sigma_{k|k-1}C'\nonumber\\
 & + \Sigma_w)^{-1}(y - C\hat{x}_{k|k-1}),\\
\Sigma_{k|k}=& \Sigma_{k|k-1} - \Sigma_{k|k-1}C'(C\Sigma_{k|k-1}C' + \Sigma_w)^{-1}C\Sigma_{k|k-1}. \nonumber
\end{eqnarray}
Time update:
\begin{eqnarray}
\hat{x}_{k+1|k} &=& A_0 + A_1 \hat{x}_{k|k},\nonumber \\
\Sigma_{k+1|k} &=& A_1\Sigma_{k|k}A'_1 + G(\hat{x}_{k|k}) \Sigma_v G(\hat{x}_{k|k}),
\end{eqnarray}
if $g^2_i(x_k), i = 1, 2, \ldots n$ is an affine function of $x_{jk}, j = 1, 2, \ldots n$.
\label{Thm:DiscreteBLUE}
\end{thm}

\begin{proof}
The Best Linear Unbiased Estimator is the basis for the Measurement Update:
$y_k = Cx_k + w_k \Rightarrow y_k|Y_{k-1} = Cx_k|Y_{k-1} + w_k$.
An affine estimator of $x_k|Y_{k-1}$ given a new measurement $y_k$ is $K_1y_k + K_2$.
For this to be unbiased, $K_2 = \hat{x}_{k|k-1} - K_1C\hat{x}_{k|k-1}$.
The affine estimator $\hat{x}_{k|k-1} + K_1(y_k - C\hat{x}_{k|k-1})$ is best in the sense of minimising the mean square error if $K_1 = \Sigma_{k|k-1}C'(C\Sigma_{k|k-1}C' + \Sigma_w)^{-1}$. 
Its covariance follows $\Sigma_{k|k} = \Sigma_{k|k-1} - \Sigma_{k|k-1}C'(C\Sigma_{k|k-1}C' + \Sigma_w)^{-1}C\Sigma_{k|k-1}$.

The Time Update follows from the properties of conditional expectations:
$x_{k+1}|Y_k = (A_0 + A_1 x_k + G(x_k)v_k)|Y_k =  A_0 + A_1 x_k|Y_k + G(x_k)v_k|Y_k = A_0 + A_1 x_k|Y_k + G(x_k)|Y_k v_k 
\Rightarrow 
\hat{x}_{k+1|k} = E\{x_{k+1}|Y_k\} = A_0 + A_1\hat{x}_{k|k}$
and
$\Sigma_{k+1|k} = E\{(x_{k+1} - x_{k+1}|Y_k)(x_{k+1} - x_{k+1}|Y_k)'|Y_k\}
= A_1\Sigma_{k|k}A'_1 + E\{G(x_k)v_kv_k'G'(x_k)|Y_k\}
= A_1\Sigma_{k|k}A'_1 + G(\hat{x}_{k|k}) \Sigma_v G(\hat{x}_{k|k})
$
if each component of $G(x_k)$ is an affine function of the components of $x_k$.
\end{proof}

We noted that this had the same form as the Kalman Filter except that the process noise covariance was updated at each time step.
While this would appear from an approximation of the system via linearization and the subsequent application of the standard Kalman Filter, as in an Extended Kalman Filter, the above theorem explicitly states the conditions under which the Kalman Filter variant is optimal.

We used a limiting process to extend the above result and obtain the optimal linear filter for continuous-discrete systems, where the process dynamics are in continuous-time and the output measurements are in discrete-time.
\begin{thm}
The Best Linear Unbiased Estimator for the continuous-discrete system
\begin{eqnarray}
dx_t &=& (A_0 + A_1x_t)dt + G(x_t)d\beta_t \\
y_k &=& Cx_{t_k} + w_k,
\end{eqnarray}
where measurements are taken at $t_k, k = 1, 2, \ldots,$, $x \in \mathbb{R}^n$, $y_k \in \mathbb{R}^m$, $\{v(t)\}$ is an $n$-vector, white noise process $v(t) \sim (0,\Sigma_v)$ with $\Sigma_v$ diagonal, $w_k \sim (0, \Sigma_w)$, and $x_0$, $\{v(t)\}$, and  $\{w_{t_k}\}$ are assumed independent, and $G(x) = diag[g_1(x), g_2(x, \ldots g_n(x)]$ is\\
Measurement update at $t = t_k$:
\begin{eqnarray}
\hat{x}_{t^+_k|t_k} =& \hat{x}_{t^-_k|t_{k-1}} + \Sigma_{t^-_k|t_{k-1}}C'(C\Sigma_{t^-_k|t_{k-1}}C' \nonumber \\
&+ \Sigma_w)^{-1}(y - C\hat{x}_{t^-_k|t_{k-1}}),\\
\Sigma_{t^+_k|t_k} =& \Sigma_{t^-_k|t_{k-1}} - \Sigma_{t^-_k|t_{k-1}}C'(C\Sigma_{t^-_k|t_{k-1}}C' \nonumber\\
&+ \Sigma_w)^{-1}C\Sigma'_{t^-_k|t_{k-1}}.\nonumber
\end{eqnarray}
Time update, $t_k \leq t < t_{k+1}$:
\begin{eqnarray}
\dot{\hat{x}}_{t|t_k} &=& A_0 + A_1 \hat{x}_{t|t_k}, \hat{x}_{t_k} = \hat{x}_{t^+_k|t_k}\\
\dot{\Sigma}_{t|t_k} &=& A_1\Sigma_{t|t_k} + \Sigma_{t|t_k}A'_1 + G(\hat{x}_{t|t_k}) \Sigma_v G(\hat{x}_{t|t_k}), \Sigma_{t_k} = \Sigma_{t^+_k|t_k} \nonumber
\end{eqnarray}
if $g^2_i(x), i = 1, 2, \ldots n$ is an affine function of $x_{j}(t), j = 1, 2, \ldots n$.
\end{thm}

\begin{proof}
The measurement update is in discrete-time and follows the Best Linear Unbiased Estimator.

The time update follows from a discretisation of the process dynamics, an application of Theorem \ref{Thm:DiscreteBLUE}, followed by a limit as the time interval of the discretization approaches zero.
Let $\Delta t$ be the time interval of discretization.
Then, for $t_k \leq t < t_{k+1}$, the discretized process dynamics are $x_{t+\Delta t}  - x_t = A_0\Delta t + A_1x_t\Delta t + \Delta t G(x_t) v_{k}$ where the covariance of $v_k$ is $\Sigma_v/\Delta t$.
The Time Update equations of Theorem \ref{Thm:DiscreteBLUE} can be applied to this discretization, $x_{t+\Delta t} = A_0\Delta t + (I + \Delta t A_1)x_t + \Delta t G(x_t) v_{k}$ to give
$\hat{x}_{t + \Delta t} = A_0\Delta t + (I + \Delta t A_1)\hat{x}_t$
and
$\Sigma_{t + \Delta t} = (I + \Delta t A_1) \Sigma_t (I + \Delta t A_1)' + \Delta t G(\hat{x}_t) (\Sigma_v/\Delta t) G(\hat{x}_t)$.
Considering the first-order terms in $\Delta t$ and taking the limit $\Delta t \rightarrow 0$ gives the desired result.
The notation $\hat{x}_{t|t_k}$ and $\Sigma_{t|t_k}$ is used to emphasize the time interval under consideration.
\end{proof}

The form of the equations were the same as the Kalman Filter for continuous-discrete systems, but with an update to the process noise covariance.
The above theorem confirmed that this variant would be optimal among all linear estimators, including the corresponding Kalman Filter.

The proof of Theorem 1 may complementarily be obtained by using a statistical approach that minimizes an objective function.
We found that the function to be minimized is a quadratic approximation to the least squares cost weighted by the noise covariances, defined recursively as
$J_{k} (z_k) = J_{k} (z_{k-1}) +  \frac{1}{2}(x_k - A_0 - A_1x_{k-1})'G^{-1}(x_{k-1})\Sigma^{-1}_vG^{-1}(x_{k-1})(x_k - A_0 - A_1x_{k-1}) + \frac{1}{2}(y_k - Cx_k)'\Sigma^{-1}_w(y_k - Cx_k)$,
where $z_k = [x_0, x_1, \ldots x_k]'$.
For completeness, this proof is placed in the appendix.

We noted that this statistical proof approach did not need the affine restriction on the process noise covariance because of the quadratic approximation to the cost function.
The affine condition was needed in the probabilistic proof approach to interchange the expectation operator and the function.
It is this advantage that allows this statistical proof approach to be seamlessly extended to the fully nonlinear setting, with nonlinear terms in the the process dynamics.

\begin{thm}
For the discrete system
\begin{eqnarray}
x_{k+1} &=& f(x_k) + G(x_k)v_k \\
y_k &=& Cx_k + w_k,
\end{eqnarray}
where $k = 0, 1, \ldots,$, $x_k \in \mathbb{R}^n$, $y_k \in \mathbb{R}^m$, $\{v_k, k =1,\ldots\}$ is an $n$-vector, white noise sequence $v_k \sim (0,\Sigma_v)$ with $\Sigma_v$ diagonal, $w_k \sim (0, \Sigma_w)$, and $x_0$, $\{v_k\}$, and  $\{w_k\}$ are assumed independent, and $G(x_k) = diag[g_1(x_k), g_2(x_k), \ldots g_n(x_k)]$,
the estimator that minimizes a quadratic approximation to a least squares cost weighted by the noise covariances,
\begin{align}
J_k(z_k) = J_{k-1}(z_{k-1}) + \frac{1}{2}(x_k - f(x_{k-1}))'G^{-1}(x_{k-1})\Sigma^{-1}_v\nonumber\\
G^{-1}(x_{k-1})(x_k - f(x_{k-1})) + \frac{1}{2}(y_k - Cx_k)'\Sigma^{-1}_w(y_k - Cx_k),
\end{align}
where $z_k = [x_0, x_1, \ldots x_k]'$, is\\
Measurement update:
\begin{eqnarray}
\hat{x}_{k|k} =& \hat{x}_{k|k-1} + \Sigma_{k|k-1}C'(C\Sigma_{k|k-1}C' + \Sigma_w)^{-1}\nonumber\\
&(y - C\Sigma_{k|k-1}),\\
\Sigma_{k|k} =& \Sigma_{k|k-1} - \Sigma_{k|k-1}C'(C\Sigma_{k|k-1}C' + \Sigma_w)^{-1}\nonumber\\
&C\Sigma_{k|k-1}.\nonumber
\end{eqnarray}
Time update:
\begin{eqnarray}
\hat{x}_{k+1|k} &=& f(\hat{x}_{k|k}),\\
\Sigma_{k+1|k} &=& Df\Sigma_{k|k}Df' + G(\hat{x}_{k|k}) \Sigma_v G(\hat{x}_{k|k}).
\end{eqnarray}
\end{thm}

\begin{proof}
The proof directly follows from the proof in the appendix.
The term corresponding to $A_1$ is replaced by the Jacobian $Df$ when the quadratic approximation to the cost is taken.

Consider a quadratic objective function for minimization,
$J_{k|k-1} (z_k) = J_{k-1}(z_{k-1}) + \frac{1}{2}(x_k -  f(x_{k-1}))'$ $G^{-1}(x_{k-1})\Sigma^{-1}_vG^{-1}(x_{k-1})(x_k -  f(x_{k-1}))$, 
where $z_k = [x_0, x_1, \ldots x_k]'$.
The weights are the inverse of the covariance functions.
The quadratic approximation to $J_{k|k-1}$, denoted $\tilde{J}_{k|k-1}$ is 
$\tilde{J}_{k|k-1}(z_k) = \tilde{J}_{k-1}(z_{k-1}) + \frac{1}{2}\tilde{r}(x_k, x_{k-1})'\Sigma^{-1}_v\tilde{r}(x_k, x_{k-1})$, 
where $\tilde{J}_{k-1}$ is the quadratic approximation to $J_{k-1}$ and 
$\tilde{r}(x_k, x_{k-1}) = G^{-1}(\hat{x}_{k-1})(x_k -  f(\hat{x}_{k-1})) + (-G^{-1}(\hat{x}_{k-1})Df+ \frac{\partial G^{-1}(x_{k-1})}{\partial x_{k-1}}(x_k -  f(\hat{x}_{k-1})))$ $(x_{k-1} - \hat{x}_{k-1})$ 
is the linear approximation to 
$r(x_k, x_{k-1}) =  G^{-1}(x_{k-1})(x_k -  f(x_{k-1}))$ around the point $(x_k, \hat{x}_{k-1})$.
The vector $z_k = [\hat{z}_{k-1},  f(\hat{x}_{k-1})]'$ sets the gradient $\nabla\tilde{J}_{k|k-1} = 0$ and gives the time update $\hat{x}_{k|k-1} =  f(\hat{x}_{k-1})$.
The Hessian $D^2\tilde{J}_{k|k-1}$ is positive definite at this point, confirming that this gives the minima.
The bottom right block of the inverse of the Hessian gives the covariance $P_{k|k-1} = (Df)P_{k-1}(Df)' + G(\hat{x}_{k-1})\Sigma_vG(\hat{x}_{k-1})$.

For the measurement update, the objective function is updated to 
$\tilde{J}_k(z_k) = \tilde{J}_{k|k-1}(z_k) + \frac{1}{2}(y_k - Cx_k)'\Sigma^{-1}_w(y_k - Cx_k)$.
The minimum of this is obtained using the Newton method with the initial guess $z_k = \hat{z}_{k|k-1}$.
As each term in this cost function is quadratic, the Newton method converges after one iteration, which gives the measurement update,
$\hat{x}_{k|k} = \hat{x}_{k|k-1} + P_{k|k}C'\Sigma_w^{-1}(y_k - C\hat{x}_{k|k-1})$.
The covariance was obtained from the bottom right block of the inverse of the Hessian $D^2\tilde{J}_k(z_k)$ as 
$P_{k|k} = P_{k|k-1} - P_{k|k-1}C'(CP_{k|k-1}C' + \Sigma_w)^{-1}CP_{k|k-1}$.
\end{proof}

We noted that this filter could be extended to nonlinear continuous-discrete dynamics using the limiting process.

\section{Example}
For completeness, the above results are illustrated through an example~\cite{dey_kalman_2019}.
Consider output data $\{y_1, y_2 \ldots y_N\}$ generated from the following discrete-time system,
\begin{eqnarray}
x_{k+1} &=& 1 + 0.99 x_k + \sqrt{100 + x_k}v_k,\\
y_k &=& x_k + w_k.
\end{eqnarray}
Here, $v_k$ and $w_k$ are independent white Gaussian processes with zero mean and unit variance ($\Sigma_v = 1 = \Sigma_w$).
The initial state is $x_1 = 1$. 
Because the process noise term depends on the state, due to the multiplication of $\sqrt{100 + x_k}$ and $v_k$ in the equation for evolution of state $x_k$, the variables $x_k$ and $y_k$ are not Gaussian.

The standard Kalman Filter in this context is\\
\textit{Measurement Update:}
\begin{eqnarray}
\hat{x}_{k|k} &=& \hat{x}_{k|k-1} + \Sigma_{k|k-1}(\Sigma_{k|k-1} + \Sigma_w)^{-1}(y_k - \hat{x}_{k|k-1}), \nonumber\\
\Sigma_{k|k} &=& \Sigma_{k|k-1} - \Sigma_{k|k-1}^2(\Sigma_{k|k-1} + \Sigma_w)^{-1}. \nonumber
\end{eqnarray}
\textit{Time Update:}
\begin{eqnarray}
\hat{x}_{k+1|k} &=& 1 + 0.99\hat{x}_{k|k},\nonumber\\
\Sigma_{k+1|k} &=& (0.99)^2\Sigma_{k|k} + \beta^2\Sigma_v. \nonumber
\end{eqnarray}
Here, $\beta$ is the process noise covariance that is a design choice.
We set $\Sigma_w = 1$ to highlight the effect of the above choice in contrast to a process noise covariance that is updated at each time step.

Based on the above results, the Kalman Filter-like algorithm with a process noise covariance update is\\
\textit{Measurement Update:}
\begin{eqnarray}
\hat{x}_{k|k} &=& \hat{x}_{k|k-1} + \Sigma_{k|k-1}(\Sigma_{k|k-1} + \Sigma_w)^{-1}(y_k - \hat{x}_{k|k-1}), \nonumber\\
\Sigma_{k|k} &=& \Sigma_{k|k-1} - \Sigma_{k|k-1}^2(\Sigma_{k|k-1} + \Sigma_w)^{-1}. \nonumber
\end{eqnarray}
\textit{Time Update:}
\begin{eqnarray}
\hat{x}_{k+1|k} &=& 1 + 0.99\hat{x}_{k|k},\nonumber\\
\Sigma_{k+1|k} &=& (0.99)^2\Sigma_{k|k} + (100 + \hat{x}_{k|k})\Sigma_v. \nonumber
\end{eqnarray}
We noted the updation of the process noise covariance term at each time step in the equation for the time update of the error covariance.

A numerical comparison of the above two filters is shown in Figure 1.
We conclude that the Kalman Filter variant that updates the process noise at each step has a better performance than a Kalman Filter with a fixed process noise covariance.

\begin{figure}[htbp]
\begin{center}
\includegraphics[scale=0.4]{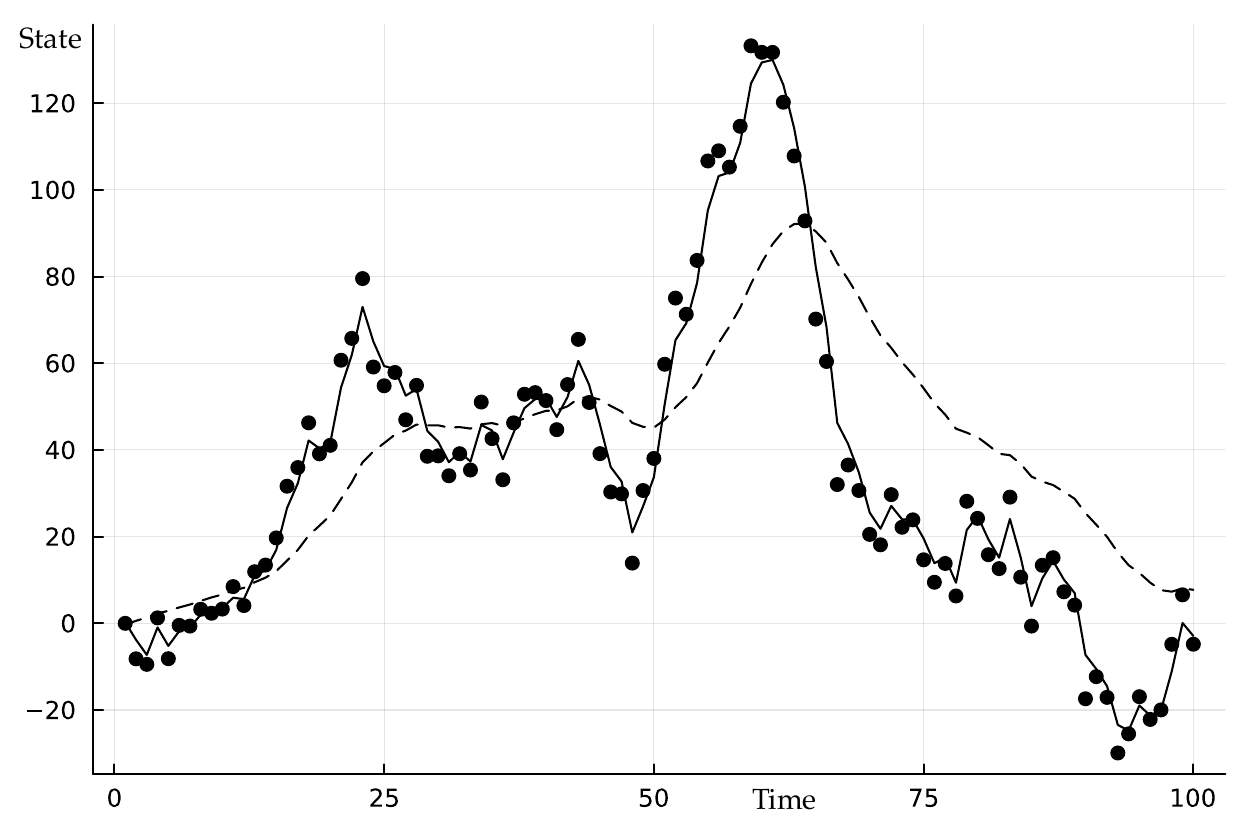}
\caption{
Numerical comparison of filters. 
Dots represent the simulated data ($N=100$).
Dashed line represents the state estimated from the Kalman Filter with $\beta = 0.1$.
Solid line represents the state estimated from a filter with process noise covariance update.
For both estimators, the initial condition $x_{1|0}$ was chosen from a Gaussian distribution with zero mean and unit variance, and $\Sigma_{1|0} = 0$.
}
\label{default}
\end{center}
\end{figure}

\section{Summary}
Stochastic models in multiple contexts have a process noise covariance that can depend on the states and the parameters.
An update of the process noise covariance at each step of the Kalman Filter yielded promising results for such models, but the theoretical justification was not completely clear.
Here we used the notion of the Best Linear Unbiased Estimator and an interpretation of the Kalman Filter as minimizing a Weighted Least Squares-like function using the Newton method to address this issue.
We proved that this algorithmic variant was the optimal linear unbiased filter for linear system dynamics when the process noice covariance depended on the state in a square root-like fashion.
We generalized this result to nonlinear systems with arbitrary process noise dependence, showing that this algorithmic variant minimized a quadratic approximation to a weighted least squares cost.
We illustrated this result using an example.
Future directions of work would be to minimize the complete weighted least squares cost, either through iterations using the Newton or Gauss-Newton methods or through methods of Interval Analysis, with the latter having multiple advantages, including of providing easily checkable criterion for solutions to not exist and of identifying multiple solutions within a given starting interval. 

\section*{Acknowledgements}
We thank Prof. Vijay Gupta for his guidance on the MMSE framework and valuable feedback.

\section*{Appendix}

\textbf{Alternate Statistical Proof of Theorem 1} 
\begin{proof}
Consider a quadratic objective function for minimization,
$J_{k|k-1} (z_k) = J_{k-1}(z_{k-1}) + \frac{1}{2}(x_k - A_0 - A_1x_{k-1})'G^{-1}(x_{k-1})\Sigma^{-1}_vG^{-1}(x_{k-1})(x_k - A_0 - A_1x_{k-1})$,
where $z_k = [x_0, x_1, \ldots x_k]'$.
The weights are the inverse of the covariance functions.
The quadratic approximation to $J_{k|k-1}$, denoted $\tilde{J}_{k|k-1}$ is 
$\tilde{J}_{k|k-1}(z_k) = \tilde{J}_{k-1}(z_{k-1}) + \frac{1}{2}\tilde{r}(x_k, x_{k-1})'\Sigma^{-1}_v\tilde{r}(x_k, x_{k-1})$, 
where $\tilde{J}_{k-1}$ is the quadratic approximation to $J_{k-1}$ and 
$\tilde{r}(x_k, x_{k-1}) = G^{-1}(\hat{x}_{k-1})(x_k - A_0 - A_1\hat{x}_{k-1}) + (-G^{-1}(\hat{x}_{k-1})A_1 + \frac{\partial G^{-1}(x_{k-1})}{\partial x_{k-1}}(x_k - A_0 - A_1\hat{x}_{k-1})) (x_{k-1} - \hat{x}_{k-1})$ 
is the linear approximation to 
$r(x_k, x_{k-1}) =  G^{-1}(x_{k-1})(x_k - A_0 - A_1x_{k-1})$ around the point $(x_k, \hat{x}_{k-1})$.
The vector $z_k = [\hat{z}_{k-1}, A_0 + A_1\hat{x}_{k-1}]'$ sets the gradient $\nabla\tilde{J}_{k|k-1} = 0$ and gives the time update $\hat{x}_{k|k-1} = A_0 + A_1\hat{x}_{k-1}$.
The Hessian $D^2\tilde{J}_{k|k-1}$ is positive definite at this point, confirming that this gives the minima.
The bottom right block of the inverse of the Hessian gives the covariance $P_{k|k-1} = A_1P_{k-1}A'_1 + G(\hat{x}_{k-1})\Sigma_vG(\hat{x}_{k-1})$.

For the measurement update, the objective function is updated to 
$\tilde{J}_k(z_k) = \tilde{J}_{k|k-1}(z_k) + \frac{1}{2}(y_k - Cx_k)'\Sigma^{-1}_w(y_k - Cx_k)$.
The minimum of this is obtained using the Newton method with the initial guess $z_k = \hat{z}_{k|k-1}$.
As each term in this cost function is quadratic, the Newton method converges after one iteration, which gives the measurement update,
$\hat{x}_{k|k} = \hat{x}_{k|k-1} + P_{k|k}C'\Sigma_w^{-1}(y_k - C\hat{x}_{k|k-1})$.
The covariance was obtained from the bottom right block of the inverse of the Hessian $D^2\tilde{J}_k(z_k)$ as 
$P_{k|k} = P_{k|k-1} - P_{k|k-1}C'(CP_{k|k-1}C' + \Sigma_w)^{-1}CP_{k|k-1}$.
\end{proof}

\bibliographystyle{unsrt}
\bibliography{KFCU}

\end{document}